%% file: main.tex
\documentclass[myepj-spec]{mySvjour}
\usepackage{graphicx}
\usepackage[pdfpagemode=UseNone]{hyperref}
\usepackage{color}
\usepackage{xspace}
\usepackage[square,sort&compress,numbers]{natbib}   
\usepackage{verbatim}
\usepackage{amsmath,amssymb,amsfonts} 
\usepackage{tabularx}
\usepackage{multicol}
\usepackage{footnote}
\usepackage{listings}
\usepackage[gen]{eurosym}
\usepackage[switch, modulo]{lineno}
\usepackage{longtable}
\usepackage[utf8]{inputenc} 
\usepackage[T1]{fontenc}    
\usepackage{hyperref}       
\usepackage{url}            
\usepackage{booktabs}       
\usepackage{multirow}
\usepackage{amsfonts}       
\usepackage{nicefrac}       
\usepackage{microtype}      
\usepackage{xcolor}         
\usepackage{graphicx}
\usepackage{float}
\usepackage{caption}
\usepackage{subcaption}
\usepackage{amsmath}
\usepackage{csquotes}
\setcitestyle{square}
\usepackage{svg}
\usepackage[title]{appendix}%

\usepackage{lmodern,bm}                
\usepackage[T1]{sansmath} 
\SetMathAlphabet{\mathsfbf}{sans}{\sansmathencoding}{\sfdefault}{bx}{sl}
\usepackage{etoolbox}
\AtBeginEnvironment{sansmath}{}{}{}

\usepackage{lipsum}

\newcommand{\pT}{\ensuremath{p_{\mathrm{T}}}}

\definecolor{darkblue1}{rgb}{0,0,.2}
\definecolor{darkblue}{rgb}{0,0,.2}
\definecolor{darkred}{rgb}{0.5,0,0}
\pagecolor{white} 
\color{black}     
\hypersetup{breaklinks=true, 
	colorlinks=true, 
	linkcolor=darkblue1, 
	menucolor=darkblue1, 
	urlcolor=darkblue1,
	citecolor=darkblue1,
	pdftitle={},
	pdfauthor={},
	pdfsubject={},
	pdfkeywords={},
	pdfproducer={}
}
\usepackage{siunitx}

\bibstyle{plain}
\begin{document}

			\begin{flushright}
				\normalsize
			\end{flushright}
			
			\vspace{-2cm}
			
			\title{\Large\boldmath On the Role of Diffractive Production in Precision Studies of W and Z Bosons at the LHC}
\author{Ynyr Harris$^1$, Matthias L. Schott$^1$, Chen Wang$^2$\footnote{corresponding author: chen.w@cern.ch}}
\institute{$^1$ Physikalisches Institut, University of Bonn, Germany, \\$^2$ Deutsches Elektronen-Synchrotron DESY, 22603 Hamburg, Germany}
			
\abstract{The increasing precision of measurements at the Large Hadron Collider (LHC) requires a detailed understanding of all contributions to electroweak boson production. In this work, we estimate the impact of single-diffractive $W$ and $Z$ boson production on high-precision observables at $\sqrt{s}=5$, $7$, $8$, and $13~\mathrm{TeV}$. The non-diffractive baseline is calculated with \textsc{DYTurbo}, while the relative single-diffractive contribution and its transverse-momentum dependence are obtained from \textsc{Herwig}. Factorisation breaking is incorporated through a $p_T$-dependent rapidity-gap survival probability calculated with the dynamic multiparton-interaction model of \textsc{Pythia8}. The diffractive Asimov spectrum is obtained by a deterministic bin-by-bin reweighting of the same high-numerical-precision \textsc{DYTurbo} cross section used for the non-diffractive baseline. With the non-perturbative parameters profiled, the nominal relative shifts in $\alpha_s$ are $-0.0072\%$, $-0.043\%$, $-0.070\%$, and $-0.067\%$ at $5$, $7$, $8$, and $13~\mathrm{TeV}$, respectively. Five Pythia8 Pomeron-flux and diffractive-PDF configurations give a maximum spread of $9.4\times10^{-6}$ in $\Delta\alpha_s$. In the $W$-mass study, the largest bias from an unmodelled survived-SD contribution is $1.58~\mathrm{MeV}$ and the largest residual after applying the corresponding model-matched template correction is $1.14~\mathrm{MeV}$ across the five survival configurations. These shifts are well below the relevant fit precision, and no phenomenologically significant impact on $m_W$ is expected within the tested models.
}

\maketitle

\section{Introduction}

Over the past decade, the physics program at the Large Hadron Collider (LHC) has undergone a clear transition from discovery-driven searches to high-precision measurements. Prominent examples include detailed studies of Higgs boson properties, as well as increasingly precise determinations of Standard Model parameters such as the top-quark mass, the $W$ boson mass \cite{CMS:2024lrd, CDF:2022hxs, ATLAS:2024erm, LHCb:2021bjt}, the effective weak mixing angle $\sin^2\theta_W$ \cite{CMS:2024ony, D0:2017ekd}, and the strong coupling constant $\alpha_s$ \cite{Camarda:2022qdg, ATLAS:2023lhg}. The precision achieved in several of these measurements has reached a level that was previously considered unattainable at hadron colliders.

Reaching such a level of precision requires not only large data samples and good control over experimental uncertainties, but also an increasingly accurate and complete theoretical description of the underlying processes. Precision measurements involving electroweak gauge bosons rely on state-of-the-art predictions based on fixed-order perturbative QCD calculations supplemented with soft-gluon resummation, and phenomenological models of the non-perturbative parts of the event as implemented in Monte Carlo (MC) event generators. These calculations provide an accurate description of key observables such as the transverse momentum ($p_T$) spectra of $W$ and $Z$ bosons, which play a central role in extractions of $\alpha_s$ and $m_W$.

However, the experimental extraction of precision observables typically assumes that vector boson production originates entirely from hard partonic scattering processes. In this approach, contributions from diffractive production mechanisms are neglected. Diffraction occurs through the exchange of a colour-singlet object known as a pomeron between the two beam particles at low momentum transfers.
Experimentally, single diffraction is characterised by the presence of a large rapidity gap (LRG, a region devoid of hadronic activity) between an intact forward beam particle and the central collision system on one side of the event. When a hard scale is present, electroweak bosons can be produced diffractively. The quantitative analysis presented below is restricted to the single-diffractive component; double-diffractive production is excluded.

The presence of diffractively produced $W$ and $Z$ bosons in experimental event samples introduces a potential source of bias.
In the single-diffractive case, since the forward proton carries away a fraction of the beam momentum, the effective centre-of-mass energy of the hard sub-process is reduced, resulting in a systematically softer boson $p_T$ spectrum compared to the non-diffractive case.
The associated rapidity gap is filled to varying degrees by soft multiple scattering between the spectator partons in the beam remnants, with a probability related to the rapidity gap survival probability, $\langle S^2\rangle$.
When the rapidity gap is preserved, the rapidity of the central hadronic system is characteristically asymmetric.
When it is destroyed by multiple scattering, the diffractive event contributes to the inclusive event sample but with an anomalous underlying event (UE) that differs from a purely non-diffractive event of the same boson $p_T$.
Since current analyses typically treat all selected events as originating from non-diffractive hard scattering, any non-negligible diffractive component may distort the inferred distributions and, consequently, the extracted physics parameters. Despite this, their impact on modern precision measurements has not been systematically quantified.

The diffractive contribution to inclusive $W/Z$ samples is observed to be at the percent level, with CDF measuring a diffractive $W$ fraction of $(1.15 \pm 0.55)\%$ \cite{CDF:1997gvd}, and CMS finding a diffractive $W$ contribution of $(50.0 \pm 9.3~(\mathrm{stat}) \pm 5.2~(\mathrm{syst}))\%$ in a large-rapidity-gap-enriched sample; this corresponds to a much smaller fraction of the fully inclusive sample \cite{Benoit:2015zmr}. Recent CDF, ATLAS, and CMS $m_W$ measurements have reached total uncertainties of about 10--16~MeV \cite{CDF:2022hxs,ATLAS:2024erm,CMS:2024lrd}. At this precision, a poorly modelled diffractive component, which distorts both the boson $\pT$ spectrum and the hadronic recoil distribution, can introduce a non-negligible systematic bias.

In this work, we present a first quantitative estimate of the impact of diffractive $W$ and $Z$ boson production on high-precision measurements at the LHC in two representative cases: the extraction of the strong coupling constant $\alpha_s$ from the $Z$ boson transverse momentum spectrum, and the determination of the $W$ boson mass. These observables are particularly sensitive to the modeling of the boson $p_T$ distribution and therefore provide an ideal testing ground for assessing potential biases. Rather than aiming for a complete theoretical description, our goal is to assess the potential size of the effect and to determine whether it needs to be taken into account in future analyses. We find small but energy-dependent deterministic shifts in $\alpha_s$. Across the five tested survival configurations, the $W$-mass bias remains below $1.58~\mathrm{MeV}$ before the template correction and below $1.14~\mathrm{MeV}$ after the corresponding model-matched correction. No phenomenologically significant impact on $m_W$ is therefore expected within these models.

The paper is organised as follows.  Section~\ref{sec:diffraction} reviews the experimental evidence for diffractive $W/Z$ production at hadron colliders and the associated phenomenology, establishing the current state of knowledge and its limitations.  In Section~\ref{sec:method}, we introduce the pseudo-dataset assembled to test the impact of a single-diffractive contribution to the electroweak boson $p_T$ spectrum based on available MC event generators.  The impact of diffraction on the extraction of $\alpha_s$ and $m_W$ is quantified in Sections~\ref{ssec:alphas} and \ref{ssec:mW}, respectively. Section~\ref{sec:conclusions} gives conclusions and outlook.

\section{Diffractive production of $W$ and $Z$ bosons at hadron colliders}\label{sec:diffraction}

Diffractive processes in hadron collisions are characterized by the exchange of a color-singlet object between the incoming hadrons, leading to final states with suppressed hadronic activity and, in some cases, large rapidity gaps. In Regge theory, such interactions are described in terms of Pomeron exchange, while in Quantum Chromodynamics (QCD) they can be interpreted as color-singlet partonic configurations. Depending on the final state, one distinguishes between single-diffractive, double-diffractive (central diffraction), and non-diffractive production mechanisms.

\begin{figure}[tbp]
    \centering
        \includegraphics[width=0.45\linewidth]{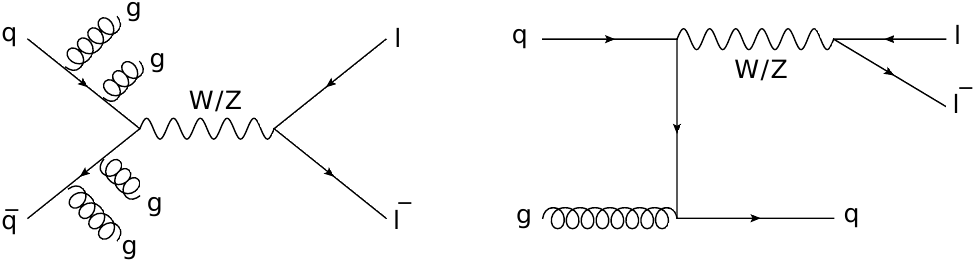}
        \hspace{0.5cm}
        \includegraphics[width=0.45\linewidth]{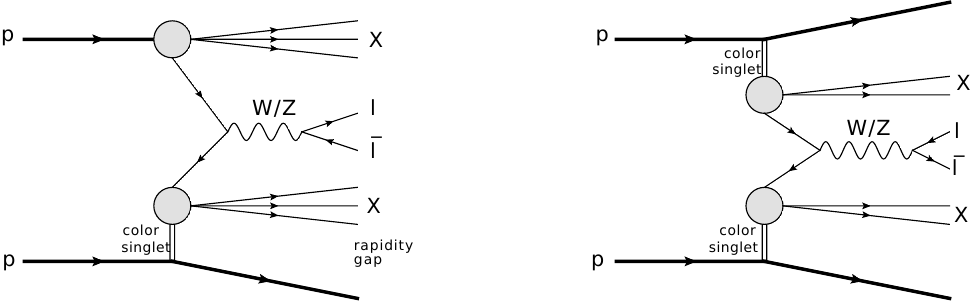}
\caption{Schematic illustration of inclusive (left) and diffractive (right) $W/Z$ boson production at hadron colliders.}
    \label{fig:WZ_schematics}
\end{figure}

Electroweak gauge bosons can be produced in diffractive interactions via partonic scattering involving constituents of the color-singlet exchange. Representative production mechanisms for inclusive and diffractive $W/Z$ boson production are illustrated in Figure~\ref{fig:WZ_schematics}. 
Inclusive production proceeds via quark--antiquark annihilation and quark--gluon interactions in perturbative QCD, where the transverse momentum of the vector boson is generated by initial-state radiation or hard parton emission. 

In contrast, diffractive production involves color-singlet exchange, leading to final states with rapidity gaps or intact protons, and results in distinct event topologies and modified kinematic distributions, particularly in the low-$p_T$ region.

Measurements at the Tevatron experiments CDF and D\O\ have established the existence of diffractive $W$ and $Z$ boson production, with cross-section fractions at the percent level. At the LHC, both the ATLAS and CMS collaborations have performed studies of diffractive and rapidity-gap-enhanced vector boson production, confirming that such processes persist at higher center-of-mass energies, albeit with significant experimental and theoretical uncertainties.

From a theoretical perspective, diffractive vector boson production poses a challenge due to the interplay between perturbative and non-perturbative QCD dynamics. In inclusive production, factorization allows for a systematic description in terms of parton distribution functions (PDFs) and perturbative matrix elements, supplemented by soft-gluon resummation to describe the low transverse momentum ($p_T$) region. In contrast, diffractive processes are commonly described using diffractive PDFs extracted from deep inelastic scattering, combined with phenomenological models of Pomeron exchange.

However, QCD factorization is known to be violated in hadron--hadron diffraction due to additional soft interactions between spectator partons, which can destroy the rapidity gap signature. This effect is typically accounted for by introducing a so-called rapidity gap survival probability, which suppresses the observable cross section relative to naive factorization-based predictions. The magnitude of this suppression depends on the collision energy and the specific process, and introduces a significant source of model dependence in theoretical predictions.

Diffractive vector-boson production is therefore commonly modeled with Monte Carlo event generators. In this analysis, the unsuppressed non-diffractive and single-diffractive boson spectra are generated with \textsc{Herwig~7.3} \cite{Bewick:2023tfi}, while the dynamic gap-survival probability is evaluated separately with \textsc{Pythia~8.3} \cite{Bierlich:2022pfr,Pythia8:hardDiff}. This treatment uses phenomenological descriptions of color-singlet exchange and multi-parton interactions, and differs fundamentally from the perturbative and resummed calculations used in precision measurements, such as those implemented in \textsc{DYTurbo} \cite{Camarda:2019zyx}, which do not include diffractive production.

Experimentally, diffractive $W$ and $Z$ boson production can be studied using several complementary techniques. One approach relies on the identification of large rapidity gaps, i.e.\ regions of the detector devoid of hadronic activity, which serve as a signature of color-singlet exchange. Alternatively, forward proton detectors allow for the direct tagging of intact scattered protons, providing a more exclusive identification of diffractive events. Both methods have been employed at the LHC, although they are subject to limitations from pile-up, detector acceptance, and background contamination from non-diffractive processes.

In the context of precision measurements, an important conceptual issue arises from the treatment of diffractive events. Standard analyses of $W$ and $Z$ boson production implicitly assume that all selected events originate from inclusive hard-scattering processes and are therefore described by perturbative QCD predictions supplemented with parton showers and non-perturbative modeling. However, the experimental event selection does not explicitly remove diffractive contributions, implying that a fraction of the selected events may stem from diffractive production.

Since diffractively produced bosons can exhibit different kinematic properties, in particular in their transverse momentum distributions, their presence can lead to distortions of observables that are crucial for precision measurements. This is especially relevant in the low-$p_T$ region, where resummation effects and non-perturbative contributions dominate, and where measurements such as the extraction of $\alpha_s$ from the $Z$ boson $p_T$ spectrum or the determination of the $W$ boson mass are particularly sensitive to the modeling of the underlying dynamics.

In the following, we aim to quantify the potential size of this effect by comparing diffractive and inclusive production models and propagating their differences to observables relevant for high-precision measurements.

\section{Methodology and Setup}\label{sec:method}

The goal of this study is to obtain a quantitative estimate of the impact of single-diffractive $W$ and $Z$ boson production on observables relevant for high-precision measurements. Both the $\alpha_s$ fits and the $W$-mass study are performed for $pp$ collisions at $\sqrt{s}=5$, $7$, $8$, and $13~\mathrm{TeV}$. We combine a high-accuracy perturbative baseline with phenomenological simulations of hard diffraction and a dynamic estimate of the rapidity-gap survival probability.

\subsection{Perturbative prediction and parameter variations}

The non-diffractive baseline is calculated with \textsc{DYTurbo}
\cite{Camarda:2019zyx}, combining the resummed and matching contributions
through order $\alpha_s^3$. We use ten $p_T$ intervals with boundaries
$0,1,2,3,4,5,7,10,15,20,$ and $25~\mathrm{GeV}$. For repeated evaluations,
the prediction in each interval $i$ is written as
\begin{equation}
 \sigma_i(g_1,q,\alpha_s)
 =R_i(g_1,q,\alpha_s)+K_i(\alpha_s),
 \label{eq:dyturbo-decomposition}
\end{equation}
where $R_i$ is the resummed component and $K_i$ is the additive non-resummed
remainder obtained from the full calculation at
$(g_1^{\mathrm{ref}},q^{\mathrm{ref}})=(0.5,0.03)$. The non-perturbative
parameters $g_1$ and $q$ affect only $R_i$, whereas $\alpha_s$ changes both
$R_i$ and $K_i$ through the perturbative coefficients and the associated PDF.
The response to these parameters is evaluated at
$\alpha_s=0.114,0.115,\ldots,0.120$ and interpolated between adjacent points.

In the $\alpha_s$ study, $g_1$, $q$, and $\alpha_s$ are profiled together
using the $Z$-boson $p_T$ spectrum. The same setup evaluates the $W$-boson
$p_T$ spectrum for the default and fitted parameter combinations; their ratio
provides the perturbative template reweighting used in the $W$-mass study.

\subsection{Diffractive modeling and reweighting}

Single-diffractive (SD) and non-diffractive (ND) vector-boson samples are
generated with a common \textsc{Herwig} setup. Their cross-section ratio
provides the relative SD normalization, while their differential distributions
provide the distinct boson-$p_T$ shape. Double-diffractive production is shown
only for completeness in Table~\ref{tab:nominal_yields} and is excluded from
the precision fits because no process-appropriate survival correction is
available.

Soft spectator interactions are accounted for with the dynamic gap-survival
model of \textsc{Pythia8} \cite{Pythia8:hardDiff}. A tentative diffractive
event survives only when no additional multiparton interaction occurs in the
proton--pomeron subsystem. In each boson-$p_T$ interval we define
\begin{equation}
 S_V^2(p_T^i)=\frac{N^{V,i}_{\mathrm{survived}}}
 {N^{V,i}_{\mathrm{tentative}}},
 \qquad V=W^+,W^-,Z.
\end{equation}
The pseudo-data spectrum is then constructed bin by bin as
\begin{equation}
 \sigma^{\mathrm{pseudo}}_{V,i}
 =\left(1+\frac{S_{V,i}^{2}\,\sigma^{\mathrm{SD}}_{V,i}}
 {\sigma^{\mathrm{ND}}_{V,i}}\right)
 \sigma^{\textsc{DYTurbo}}_{V,i}.
 \label{eq:diff-correction}
\end{equation}
Thus \textsc{Herwig} supplies the SD-to-ND normalization and shape, while
\textsc{Pythia8} supplies the process- and energy-dependent survival factor.
The latter is always applied bin by bin rather than as a constant.

The nominal \textsc{Pythia8} configuration combines the MBR Pomeron flux with
the H1 2006 Fit B LO DPDF \cite{Ciesielski:2012mc, H1:2006zyl}. Four
alternatives use matched H1 2006 Fit A or Fit B fluxes and NLO DPDFs, or the
GKG18 Fit A or Fit B LO DPDFs \cite{Goharipour:2018yov}. For the GKG18 cases,
the H1-like flux parameters are
$(\epsilon,\alpha'_{\mathbb{P}},B_0)=(0.0938,0,7.0~\mathrm{GeV}^{-2})$ and
$(0.0988,0,7.0~\mathrm{GeV}^{-2})$, respectively. The five configurations are
treated as alternative models.

\begin{figure}[htbp]
    \centering
    \begin{subfigure}{0.32\textwidth}
        \centering
        \includegraphics[width=\linewidth]{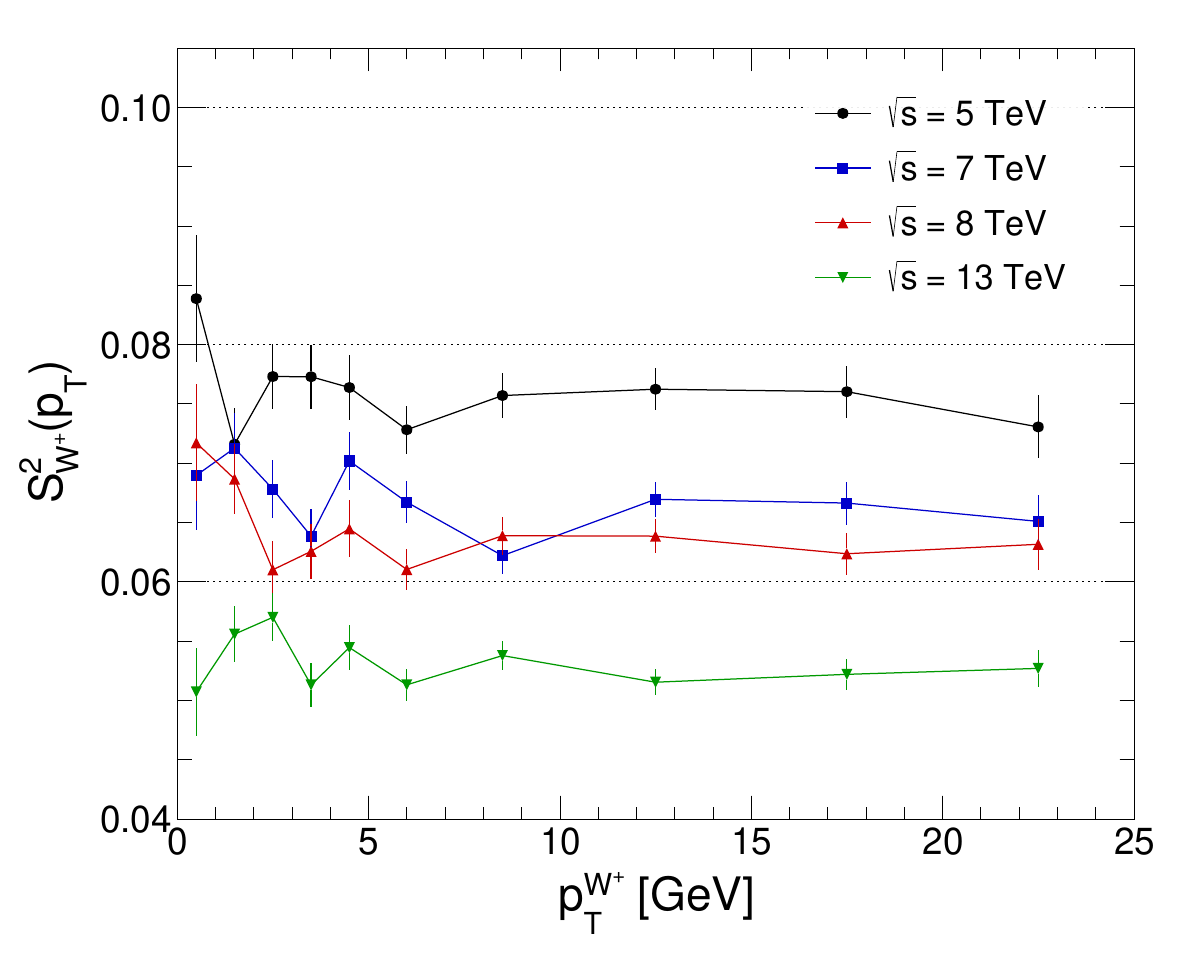}
        \caption{$W^+$ production}
    \end{subfigure}
    \hfill
    \begin{subfigure}{0.32\textwidth}
        \centering
        \includegraphics[width=\linewidth]{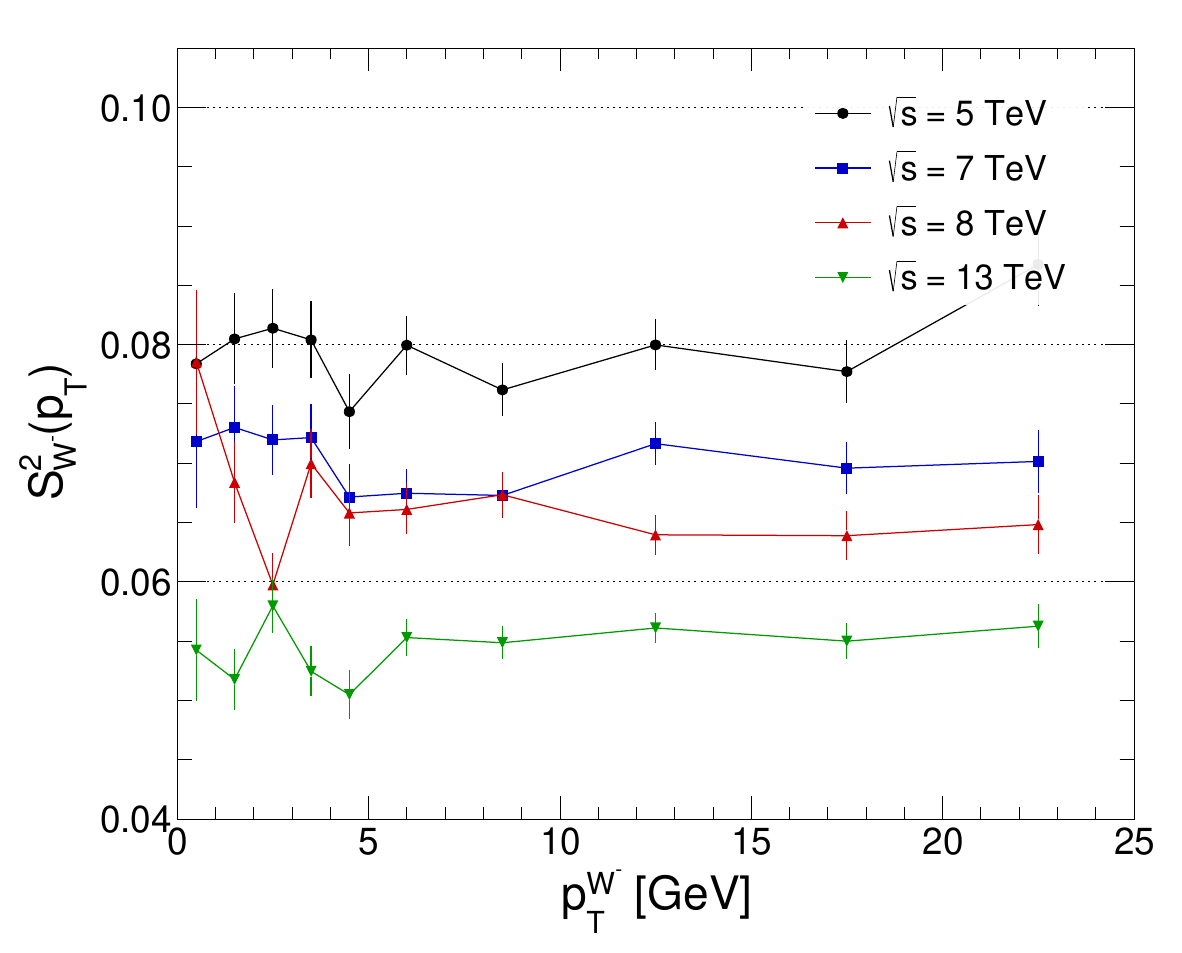}
        \caption{$W^-$ production}
    \end{subfigure}
    \hfill
    \begin{subfigure}{0.32\textwidth}
        \centering
        \includegraphics[width=\linewidth]{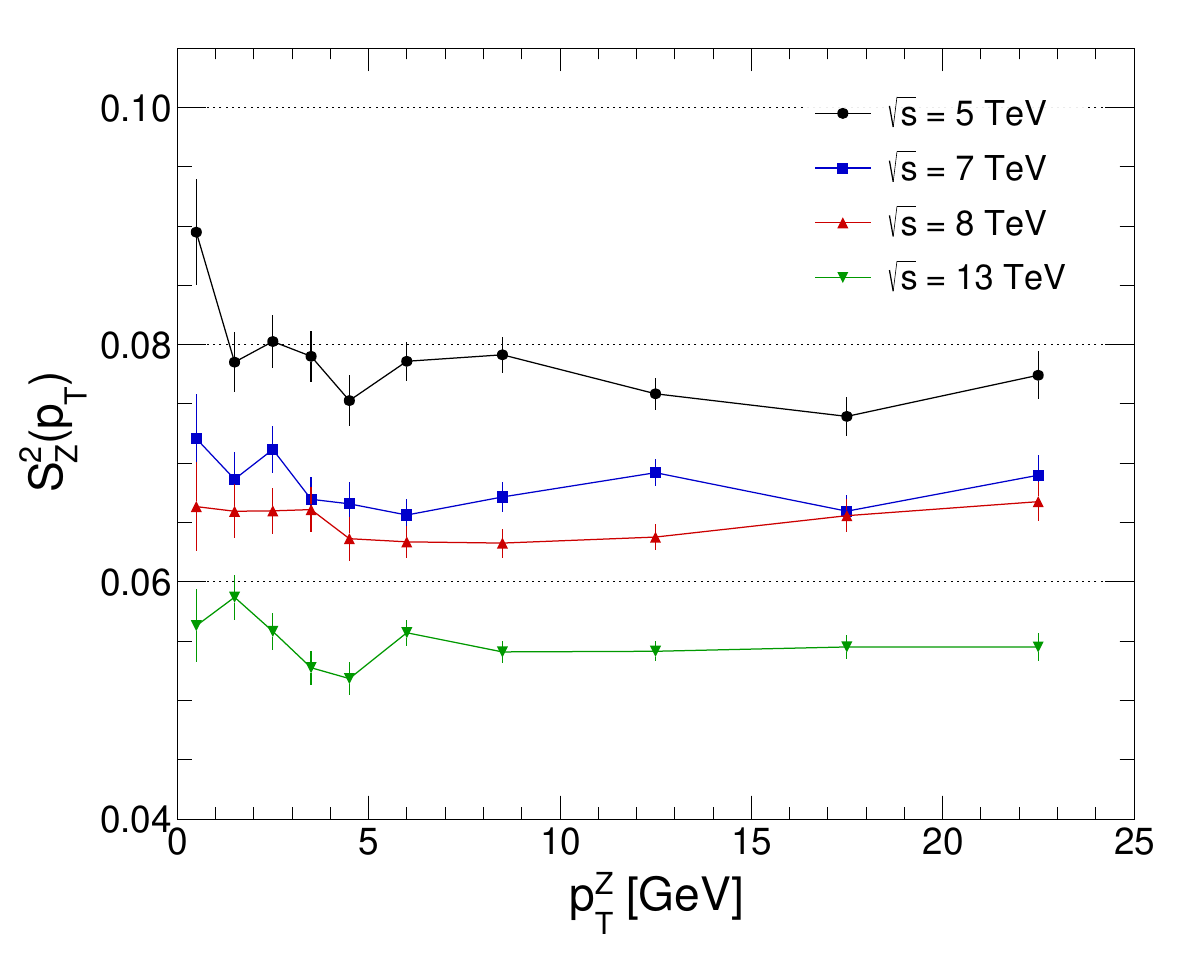}
        \caption{$Z$ production}
    \end{subfigure}
    \caption{Nominal Pythia8 dynamic gap-survival probabilities as functions
    of vector-boson transverse momentum at $\sqrt{s}=5$, $7$, $8$, and
    $13~\mathrm{TeV}$. The MBR Pomeron flux and H1 2006 Fit B LO DPDF are
    used in all panels. Vertical error bars show the finite-simulation
    statistical uncertainties.}
    \label{fig:pythia-survival}
\end{figure}

The unsuppressed \textsc{Herwig} cross sections are listed in
Table~\ref{tab:nominal_yields}. Table~\ref{tab:diffraction_fractions}
summarizes the nominal integrated survival probabilities and the corresponding
effective survived-SD fractions, while Figure~\ref{fig:pythia-survival} shows
their $p_T$ dependence. Each simulation is continued until every populated
$p_T$ bin reaches an $8\%$ relative statistical precision; all differential
fits retain the full $S_V^2(p_T)$ dependence.

\input{xsecs}

For the integrated fractions in Table~\ref{tab:diffraction_fractions},
$S_V^2$ is weighted by the corresponding Herwig SD spectrum over
$0<p_T^V<25~\mathrm{GeV}$; the $W^+$ and $W^-$ contributions are weighted
separately.

\begin{table}[htbp]
\centering
\footnotesize
\setlength{\tabcolsep}{3.5pt}
\begin{tabular}{c|ccc|cccc}
\hline
& \multicolumn{3}{c|}{Integrated survival probability}
& \multicolumn{4}{c}{Effective survived-SD fraction [\%]} \\
$\sqrt{s}$
& $\langle S_{W^+}^{2}\rangle$
& $\langle S_{W^-}^{2}\rangle$
& $\langle S_Z^{2}\rangle$
& $W$ full & $W$ fid. & $Z$ full & $Z$ fid. \\
\hline
$5~\mathrm{TeV}$
& $0.0755\pm0.0008$ & $0.0793\pm0.0009$ & $0.0776\pm0.0006$
& 0.179 & 0.171 & 0.226 & 0.0842 \\
$7~\mathrm{TeV}$
& $0.0663\pm0.0007$ & $0.0698\pm0.0008$ & $0.0677\pm0.0005$
& 0.144 & 0.134 & 0.142 & 0.0518 \\
$8~\mathrm{TeV}$
& $0.0634\pm0.0007$ & $0.0656\pm0.0008$ & $0.0646\pm0.0005$
& 0.140 & 0.128 & 0.154 & 0.0549 \\
$13~\mathrm{TeV}$
& $0.0528\pm0.0005$ & $0.0549\pm0.0006$ & $0.0545\pm0.0004$
& 0.101 & 0.0886 & 0.141 & 0.0495 \\
\hline
\end{tabular}
\caption{Nominal Pythia8 dynamic gap-survival probabilities and the
corresponding effective survived-SD fractions for the MBR Pomeron flux and H1
2006 Fit B LO DPDF configuration. The survival probabilities are integrated
over $0<p_T^V<25~\mathrm{GeV}$; their uncertainties are statistical and arise
from the finite tentative and survived event samples. The fractions give the
survived-SD contribution to the sum of ND and survived-SD production. DD
production is excluded.}
\label{tab:diffraction_fractions}
\end{table}

\begin{figure}[H]
    \centering
    \begin{subfigure}{0.48\textwidth}
        \centering
        \includegraphics[width=\linewidth]{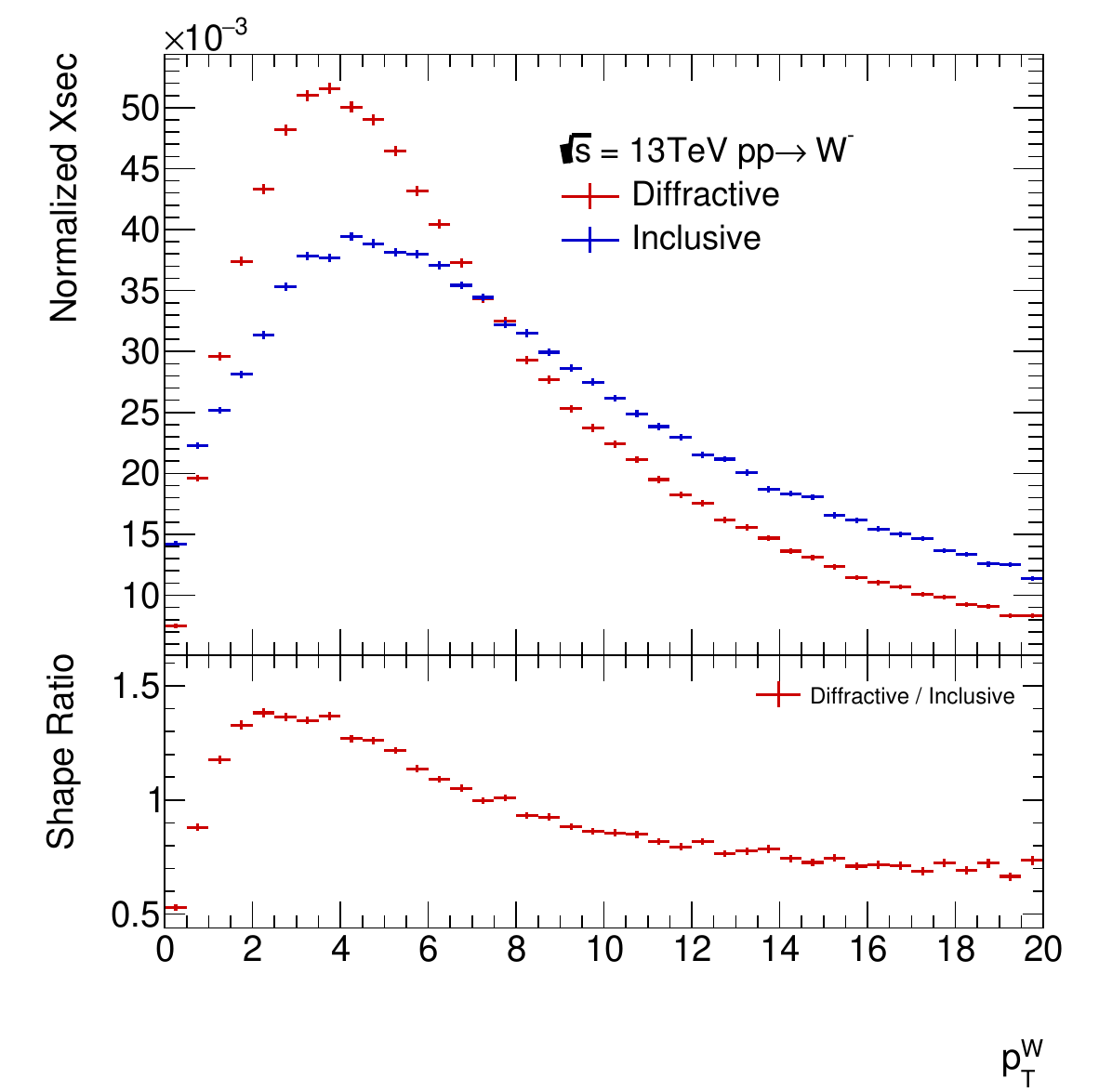}
        \caption{$W^-$ production}
    \end{subfigure}
    \hfill
    \begin{subfigure}{0.48\textwidth}
        \centering
        \includegraphics[width=\linewidth]{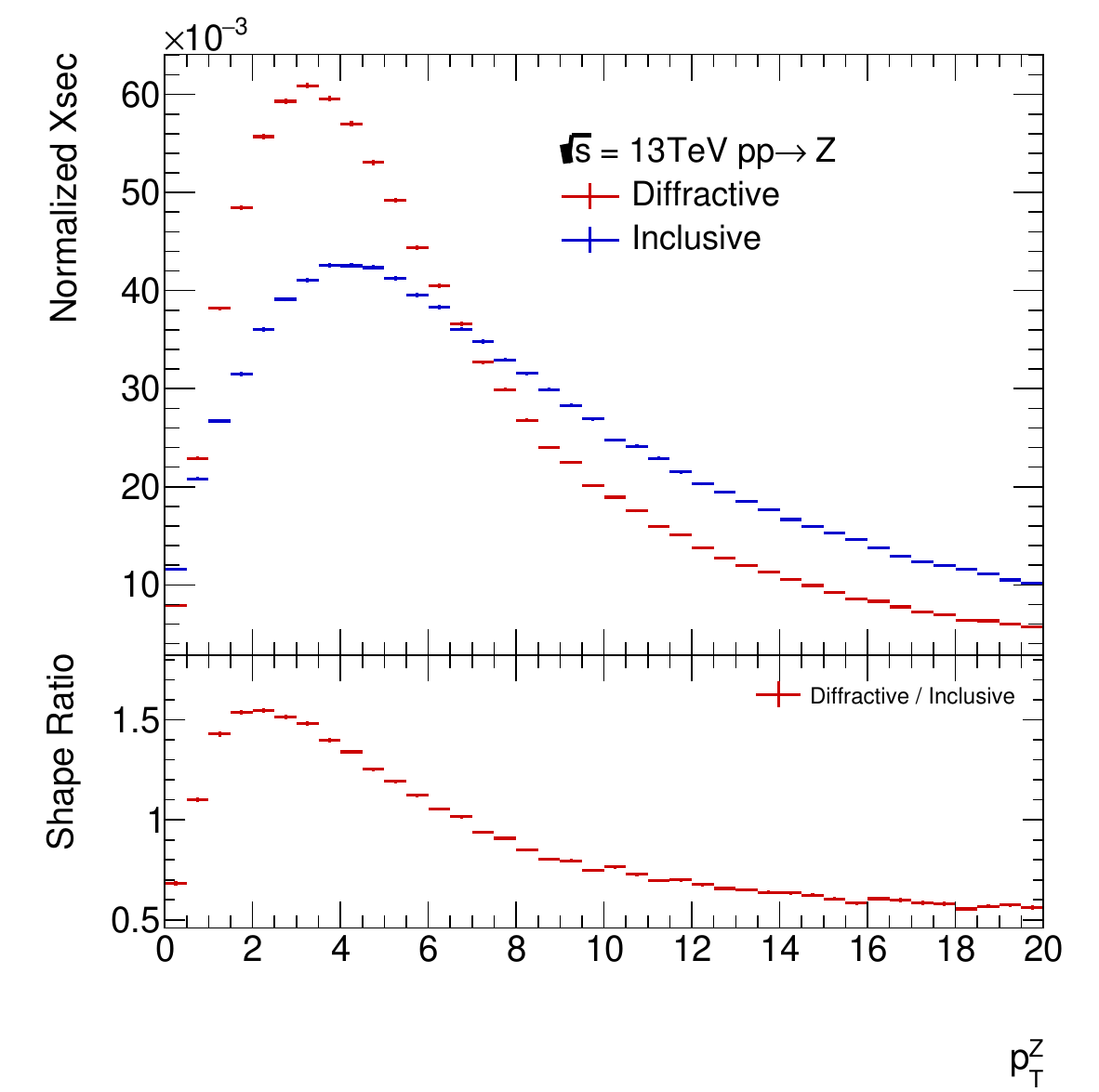}
        \caption{$Z$ production}
    \end{subfigure}
    \caption{Comparison of the normalized transverse-momentum distributions
    for diffractive and non-diffractive $W^-$ and $Z$ production at
    $\sqrt{s}=13~\mathrm{TeV}$ in the full phase space. The lower panels show
    the diffractive-to-non-diffractive shape ratios. All distributions are
    normalized to unity to expose the intrinsic difference between the two
    production mechanisms.}
    \label{fig:pt-shapes}
\end{figure}

Figure~\ref{fig:pt-shapes} compares representative normalized Herwig
diffractive and non-diffractive boson-$p_T$ spectra at
$\sqrt{s}=13~\mathrm{TeV}$ in the full phase space.

\clearpage

\subsection{Analysis phase spaces}

The $\alpha_s$ determination uses the full-leptonic-phase-space $Z$-boson
$p_T$ distribution, with no charged-lepton $p_T$ or pseudorapidity
requirements, and is fitted over $0<p_T^Z<25~\mathrm{GeV}$. The $W$-mass
study instead uses $pp\to W\to e\nu$ events satisfying
$p_T^W<25~\mathrm{GeV}$, $p_T^e>25~\mathrm{GeV}$,
$p_T^\nu>25~\mathrm{GeV}$, and $m_T>50~\mathrm{GeV}$.
Tables~\ref{tab:nominal_yields} and~\ref{tab:diffraction_fractions} also report
complementary full and fiducial quantities, while
Figure~\ref{fig:pt-shapes} uses the full phase space solely to illustrate the
intrinsic boson-$p_T$ shape difference.

\section{Results}

In this section, we estimate the impact of diffractive $W$ and $Z$ boson production on two representative high-precision measurements: the determination of the strong coupling constant $\alpha_s$ and the measurement of the $W$ boson mass. These observables are particularly sensitive to the modeling of the vector boson transverse momentum spectrum and therefore provide an ideal testing ground for assessing potential biases from neglected diffractive contributions.

\subsection{Impact on the determination of $\alpha_s$}\label{ssec:alphas}

We use normalized shape fits to compare the ND reference with the survived-SD
pseudo-data of Eq.~\eqref{eq:diff-correction} at each collision energy. The
parameters $g_1$, $q$, and $\alpha_s$ are free in both fits, with the
prediction evaluated using the decomposition in
Eq.~\eqref{eq:dyturbo-decomposition}.

The quantity of interest is the deterministic displacement between the two strongly correlated Asimov fits,
\begin{equation}
 \Delta\alpha_s^{\mathrm{diff}}
 =\alpha_s^{\mathrm{fit,diff}}-\alpha_s^{\mathrm{fit,ND}}.
 \label{eq:delta-alpha-diff}
\end{equation}
Because the two spectra use the same high-precision \textsc{DYTurbo} baseline,
its numerical uncertainty is strongly correlated and largely cancels in
Eq.~\eqref{eq:delta-alpha-diff}. The finite \textsc{Pythia8} uncertainty is
propagated by shifting each $S_Z^2(p_T)$ bin up and down, refitting all three
parameters, and adding the resulting half-differences in quadrature. PDF,
scale, generator, and overall survival-model uncertainties are not included.

The ND fits recover the common input $\alpha_s=0.118$ at all four energies.
The corresponding deterministic shifts are summarized in
Table~\ref{tab:alphas-models}.

\begin{table*}[htbp]
\centering
\small
\setlength{\tabcolsep}{6pt}
\caption{Dependence of the deterministic $\alpha_s$ displacement on the
\textsc{Pythia8} Pomeron-flux and DPDF configuration. Entries are
$10^5\Delta\alpha_s^{\mathrm{diff}}\pm10^5\sigma_{S^2\mathrm{\,stat}}$;
the first row is the nominal configuration. The spread among rows is a model
envelope, not a complete theoretical uncertainty.}
\label{tab:alphas-models}
\begin{tabular}{lcccc}
\hline
Pythia8 configuration & $5~\mathrm{TeV}$ & $7~\mathrm{TeV}$ & $8~\mathrm{TeV}$ & $13~\mathrm{TeV}$ \\
\hline
MBR / H1 Fit B LO (nominal) & $-0.844\pm0.048$ & $-5.06\pm0.20$ & $-8.26\pm0.27$ & $-7.94\pm0.24$ \\
H1 Fit A / H1 Fit A NLO     & $-1.000\pm0.050$ & $-5.17\pm0.21$ & $-9.20\pm0.27$ & $-7.99\pm0.25$ \\
H1 Fit B / H1 Fit B NLO     & $-1.042\pm0.052$ & $-5.18\pm0.21$ & $-8.39\pm0.29$ & $-7.92\pm0.27$ \\
GKG18 Fit A LO              & $-1.111\pm0.048$ & $-4.93\pm0.22$ & $-8.50\pm0.27$ & $-7.92\pm0.28$ \\
GKG18 Fit B LO              & $-1.052\pm0.052$ & $-5.14\pm0.22$ & $-8.64\pm0.29$ & $-8.22\pm0.25$ \\
\hline
\end{tabular}
\end{table*}

All five configurations give a small negative displacement. In the nominal
case the largest is $-8.26\times10^{-5}$ ($-0.070\%$) at $8~\mathrm{TeV}$,
where the largest full model spread, $9.4\times10^{-6}$, is also found. This
envelope does not cover the unknown overall survival normalization or
differences between generators.

\subsection{Impact on the $W$ boson mass}\label{ssec:mW}

\begin{figure}[H]
    \centering
    \begin{subfigure}{0.48\textwidth}
        \centering
        \includegraphics[width=\linewidth]{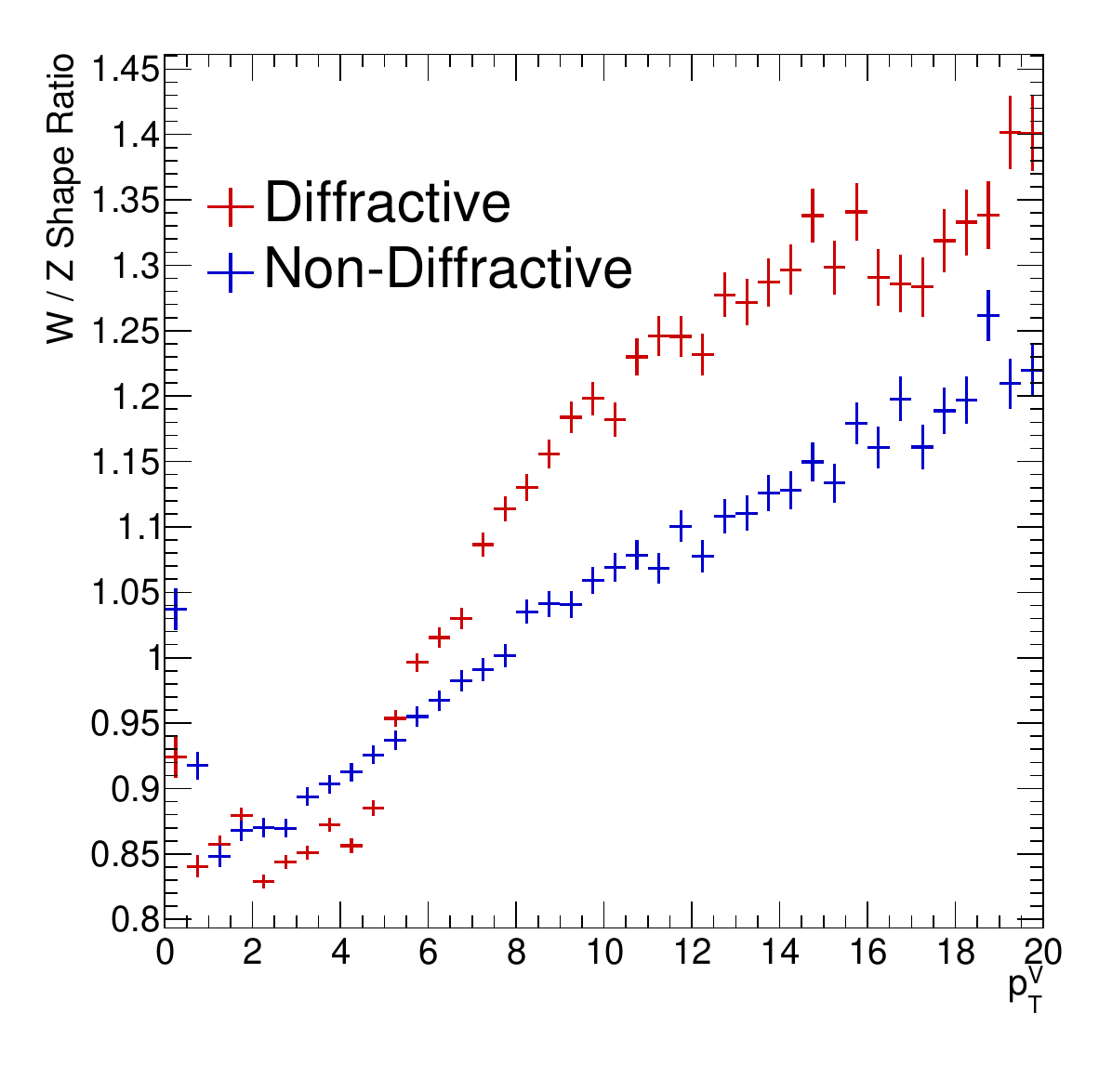}
        \caption{$W/Z$ shape ratios for survived SD and ND production}
    \end{subfigure}
    \hfill
    \begin{subfigure}{0.48\textwidth}
        \centering
        \includegraphics[width=\linewidth]{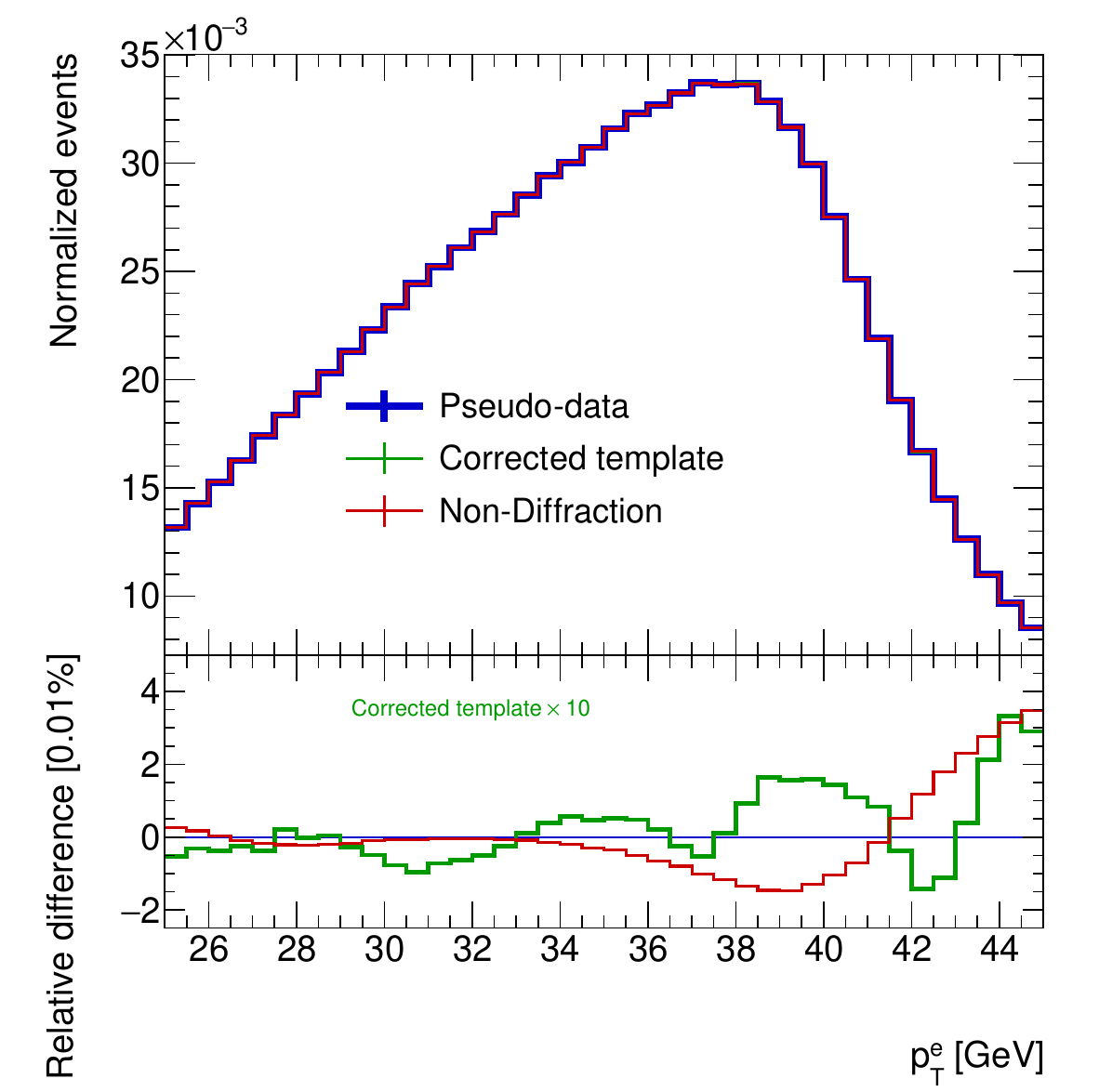}
        \caption{Charged-lepton $p_T$ templates}
    \end{subfigure}
    \caption{Impact of survived single-diffractive production on the modeling of the $W$-boson transverse momentum and decay kinematics. The left panel compares the normalized $W/Z$ shape ratios for the survived-SD and ND components. The upper-right panel shows the normalized charged-lepton $p_T$ spectrum for non-diffractive production, the survived-SD pseudo-data, and the DYTurbo-corrected template over $25<p_T^e<45~\mathrm{GeV}$. The lower-right panel shows $100\,(\mathrm{template}-\mathrm{pseudo\text{-}data})/\mathrm{pseudo\text{-}data}$; the corrected-template residual (green) is multiplied by 10 for visibility, and the blue line marks zero.}
    \label{fig:mw}
\end{figure}

The $W$-mass sensitivity is evaluated with samples of ten million \textsc{Pythia8} $pp\to W\to e\nu$ events at each of $\sqrt{s}=5$, $7$, $8$, and $13~\mathrm{TeV}$. We require $p_T^W<25~\mathrm{GeV}$, $p_T^e>25~\mathrm{GeV}$, $p_T^\nu>25~\mathrm{GeV}$, and $m_T>50~\mathrm{GeV}$. The pseudo-data are reweighted event by event with the charge-dependent factor $(\mathrm{ND}+S^2(p_T^W)\,\mathrm{SD})/\mathrm{ND}$, using the Herwig SD and ND spectra and the Pythia gap-survival probability; DD production is excluded. Figure~\ref{fig:mw} shows the resulting normalized shape distortion and the corresponding DYTurbo template correction for the representative $8~\mathrm{TeV}$ sample. The fitted mass displacements are summarized in Table~\ref{tab:mw-models}.

\clearpage

\begin{table*}[p]
\centering
\setlength{\tabcolsep}{3pt}
\renewcommand{\arraystretch}{1.08}
\caption{Dependence of the fitted $W$-mass displacement on the Pythia8 survival configuration. All entries are in MeV. The raw displacement, $\Delta m_W^{\mathrm{raw}}=m_W^{\mathrm{diff}}-m_W^{\mathrm{ND}}$, is obtained when survived-SD production is included only in the pseudo-data. The residual, $\Delta m_W^{\mathrm{res}}=m_W^{\mathrm{corrected}}-m_W^{\mathrm{ND}}$, is obtained after applying the model-matched DYTurbo correction to the templates. The same Pythia8 survival model is used for the pseudo-data and for the corresponding $Z$-boson fit from which the template correction is obtained. The quoted uncertainties arise only from the finite Pythia8 samples used to determine $S^2(p_T^W)$; the strongly correlated single-fit statistical uncertainties are not added in quadrature.}

\label{tab:mw-models}
\resizebox{\textwidth}{!}{%
\begin{tabular}{cllcccc}
\hline
Observable & Pythia8 configuration & Displacement & $5~\mathrm{TeV}$ & $7~\mathrm{TeV}$ & $8~\mathrm{TeV}$ & $13~\mathrm{TeV}$ \\
\hline
\multirow{10}{*}{$m_T$}
 & \multirow{2}{*}{MBR / H1 Fit B LO (nominal)} & Raw      & $-0.0405\pm0.0011$ & $-0.0349\pm0.0009$ & $-0.0339\pm0.0009$ & $-0.0225\pm0.0006$ \\
 &                                               & Residual & $-0.0095\pm0.0011$ & $-0.0049\pm0.0009$ & $ 0.0040\pm0.0009$ & $ 0.0195\pm0.0006$ \\
 & \multirow{2}{*}{H1 Fit A / H1 Fit A NLO}      & Raw      & $-0.0396\pm0.0011$ & $-0.0341\pm0.0009$ & $-0.0331\pm0.0009$ & $-0.0219\pm0.0006$ \\
 &                                               & Residual & $-0.0111\pm0.0011$ & $-0.0047\pm0.0009$ & $ 0.0068\pm0.0009$ & $ 0.0077\pm0.0006$ \\
 & \multirow{2}{*}{H1 Fit B / H1 Fit B NLO}      & Raw      & $-0.0426\pm0.0011$ & $-0.0357\pm0.0009$ & $-0.0336\pm0.0009$ & $-0.0226\pm0.0007$ \\
 &                                               & Residual & $-0.0127\pm0.0011$ & $-0.0041\pm0.0009$ & $ 0.0067\pm0.0009$ & $ 0.0033\pm0.0007$ \\
 & \multirow{2}{*}{GKG18 Fit A LO}               & Raw      & $-0.0418\pm0.0011$ & $-0.0350\pm0.0010$ & $-0.0343\pm0.0009$ & $-0.0228\pm0.0007$ \\
 &                                               & Residual & $-0.0119\pm0.0011$ & $-0.0038\pm0.0010$ & $ 0.0080\pm0.0009$ & $ 0.0227\pm0.0007$ \\
 & \multirow{2}{*}{GKG18 Fit B LO}               & Raw      & $-0.0399\pm0.0012$ & $-0.0318\pm0.0009$ & $-0.0343\pm0.0009$ & $-0.0229\pm0.0006$ \\
 &                                               & Residual & $-0.0096\pm0.0012$ & $ 0.0015\pm0.0009$ & $ 0.0096\pm0.0009$ & $ 0.0060\pm0.0006$ \\
\hline
\multirow{10}{*}{$p_T^e$}
 & \multirow{2}{*}{MBR / H1 Fit B LO (nominal)} & Raw      & $-1.496\pm0.039$ & $-1.239\pm0.031$ & $-1.228\pm0.032$ & $-0.833\pm0.022$ \\
 &                                               & Residual & $-0.408\pm0.039$ & $-0.241\pm0.031$ & $ 0.060\pm0.032$ & $ 0.733\pm0.022$ \\
 & \multirow{2}{*}{H1 Fit A / H1 Fit A NLO}      & Raw      & $-1.477\pm0.039$ & $-1.210\pm0.031$ & $-1.195\pm0.031$ & $-0.821\pm0.022$ \\
 &                                               & Residual & $-0.491\pm0.039$ & $-0.243\pm0.031$ & $ 0.155\pm0.031$ & $ 0.244\pm0.022$ \\
 & \multirow{2}{*}{H1 Fit B / H1 Fit B NLO}      & Raw      & $-1.581\pm0.040$ & $-1.271\pm0.032$ & $-1.215\pm0.032$ & $-0.840\pm0.023$ \\
 &                                               & Residual & $-0.560\pm0.040$ & $-0.239\pm0.032$ & $ 0.147\pm0.032$ & $ 0.224\pm0.023$ \\
 & \multirow{2}{*}{GKG18 Fit A LO}               & Raw      & $-1.533\pm0.039$ & $-1.252\pm0.032$ & $-1.238\pm0.030$ & $-0.850\pm0.024$ \\
 &                                               & Residual & $-0.525\pm0.039$ & $-0.239\pm0.032$ & $ 0.194\pm0.030$ & $ 1.005\pm0.025$ \\
 & \multirow{2}{*}{GKG18 Fit B LO}               & Raw      & $-1.472\pm0.041$ & $-1.134\pm0.031$ & $-1.221\pm0.032$ & $-0.862\pm0.022$ \\
 &                                               & Residual & $-0.460\pm0.041$ & $-0.046\pm0.031$ & $ 0.273\pm0.032$ & $ 0.255\pm0.022$ \\
\hline
\multirow{10}{*}{$p_T^\nu$}
 & \multirow{2}{*}{MBR / H1 Fit B LO (nominal)} & Raw      & $-1.472\pm0.041$ & $-1.236\pm0.034$ & $-1.175\pm0.034$ & $-0.804\pm0.023$ \\
 &                                               & Residual & $-0.293\pm0.041$ & $-0.154\pm0.034$ & $ 0.170\pm0.034$ & $ 0.810\pm0.023$ \\
 & \multirow{2}{*}{H1 Fit A / H1 Fit A NLO}      & Raw      & $-1.421\pm0.042$ & $-1.210\pm0.034$ & $-1.144\pm0.032$ & $-0.782\pm0.022$ \\
 &                                               & Residual & $-0.341\pm0.042$ & $-0.152\pm0.034$ & $ 0.270\pm0.032$ & $ 0.202\pm0.022$ \\
 & \multirow{2}{*}{H1 Fit B / H1 Fit B NLO}      & Raw      & $-1.546\pm0.043$ & $-1.262\pm0.034$ & $-1.164\pm0.033$ & $-0.811\pm0.024$ \\
 &                                               & Residual & $-0.415\pm0.043$ & $-0.132\pm0.034$ & $ 0.263\pm0.033$ & $ 0.214\pm0.024$ \\
 & \multirow{2}{*}{GKG18 Fit A LO}               & Raw      & $-1.513\pm0.042$ & $-1.234\pm0.035$ & $-1.194\pm0.032$ & $-0.813\pm0.025$ \\
 &                                               & Residual & $-0.384\pm0.042$ & $-0.123\pm0.035$ & $ 0.314\pm0.032$ & $ 1.138\pm0.025$ \\
 & \multirow{2}{*}{GKG18 Fit B LO}               & Raw      & $-1.428\pm0.044$ & $-1.120\pm0.034$ & $-1.204\pm0.034$ & $-0.804\pm0.022$ \\
 &                                               & Residual & $-0.280\pm0.044$ & $ 0.059\pm0.034$ & $ 0.375\pm0.034$ & $ 0.185\pm0.022$ \\
\hline
\end{tabular}
}
\end{table*}

\clearpage

Mass templates are obtained by Breit--Wigner reweighting and fitted separately to the $m_T$, charged-lepton $p_T$, and neutrino $p_T$ distributions. We compare a nominal fit, a fit in which survived-SD production is present only in the pseudo-data, and a corrected fit in which the DYTurbo correction inferred from the corresponding $Z$-boson fit is also applied to the templates. For each of the five Pythia8 configurations, the same survival model is used both to construct the pseudo-data and to obtain the $Z$-fit parameters transferred to the $W$ templates. The same generated events enter all three fits, so the relevant quantities are their strongly correlated mass displacements.

Figure~\ref{fig:mw} and Table~\ref{tab:mw-models} show that the effect is small. Across all energies, observables, and survival configurations, the largest bias obtained when survived-SD production is omitted from the templates is $1.581\pm0.040~\mathrm{MeV}$, while the largest residual after applying the corresponding model-matched correction is $1.138\pm0.025~\mathrm{MeV}$. These extrema occur in the $5~\mathrm{TeV}$ H1 Fit B charged-lepton $p_T$ fit and the $13~\mathrm{TeV}$ GKG18 Fit A neutrino $p_T$ fit, respectively. The largest model spread of the corrected residual at fixed energy and observable is $0.953~\mathrm{MeV}$. The transverse-mass fits are essentially unaffected: their largest raw displacement is $0.043~\mathrm{MeV}$ and their largest corrected residual is $0.023~\mathrm{MeV}$. The largest propagated uncertainty from the finite Pythia8 survival sample is $0.044~\mathrm{MeV}$. These values are well below both the $3$--$10~\mathrm{MeV}$ single-fit precision of the simulated samples and the uncertainties of current experimental measurements. No phenomenologically significant impact on $m_W$ is therefore expected within the tested models.

\section{Conclusion}\label{sec:conclusions}

We have quantified the impact of diffractive $W$ and $Z$ production on the extraction of $\alpha_s$ from the low-$p_T^Z$ spectrum and on the determination of $m_W$. The calculation combines Herwig SD shapes, bin-dependent Pythia8 gap-survival probabilities, and high-precision \textsc{DYTurbo} spectra. The nominal prediction retains survived SD production and excludes DD production, while five Pythia8 flux--DPDF configurations are used to probe the modeling dependence.

The nominal $\alpha_s$ displacements are of order $10^{-5}$, and the largest spread among the tested survival configurations is $9.4\times10^{-6}$. For the $W$-mass fits, the largest bias before correcting the templates is $1.6~\mathrm{MeV}$, and all residuals after the model-matched correction remain below $1.1~\mathrm{MeV}$; the transverse-mass fits are essentially unaffected. These shifts are below both the statistical precision of the simulated fits and that of current experimental measurements, so no phenomenologically significant impact on $m_W$ is expected within the tested models.

These results should be regarded as a first estimate. Varying the Pomeron flux and the DPDF does not capture the full uncertainty on the gap-survival probability, which depends on both the process and the scale. Dedicated measurements of the $Z$-boson transverse-momentum spectrum in diffraction-enriched samples are needed to constrain the survival probability, validate the modelling, and control this contribution in future precision measurements.

{\small
\bibliographystyle{unsrt}
\bibliography{./Bibliography}
}

\end{document}

%% file: xsecs.tex
\begin{table}[t]
\centering
\begin{tabular}{cccccc}
\toprule
 $\sigma$ [pb] & & 5 TeV ($pp$) & 7 TeV ($pp$) & 8 TeV ($pp$) & 13 TeV ($pp$) \\
\midrule
\multirow{8}{*}{\textbf{$Z \rightarrow \mu \mu$}} & & \multicolumn{4}{c}{Inclusive} \\\cline{2-6}
 & $\sigma_\text{ND}$ & $853 \pm 4$ & $1214 \pm 5$ & $1393 \pm 6$ & $2254 \pm 10$ \\
 & $\sigma_\text{SD}$ & $22.97 \pm 0.16$ & $23.45 \pm 0.15$ & $30.47 \pm 0.20$ & $52.19 \pm 0.34$ \\
 & $\sigma_\text{DD}$ & $0.789 \pm 0.005$ & $1.138 \pm 0.008$ & $0.986 \pm 0.007$ & $1.745 \pm 0.012$ \\\cline{2-6}
 & & \multicolumn{4}{c}{Fiducial} \\\cline{2-6}
 & $\sigma_\text{ND}^\text{fid.}$ & $226.1 \pm 1.1$ & $302.5 \pm 1.4$ & $339.5 \pm 1.6$ & $515.0 \pm 2.5$ \\
 & $\sigma_\text{SD}^\text{fid.}$ & $2.291 \pm 0.017$ & $2.170 \pm 0.015$ & $2.697 \pm 0.020$ & $4.322 \pm 0.032$ \\
 & $\sigma_\text{DD}^\text{fid.}$ & $0.0471 \pm 0.0004$ & $0.0656 \pm 0.0005$ & $0.0546 \pm 0.0004$ & $0.0913 \pm 0.0007$ \\
 \midrule
 
\multirow{8}{*}{\textbf{$W \rightarrow \mu \nu$}} & & \multicolumn{4}{c}{Inclusive} \\\cline{2-6}
 & $\sigma_\text{ND}$ & $5184 \pm 23$ & $7567 \pm 33$ & $8710 \pm 40$ & $14770 \pm 60$ \\
 & $\sigma_\text{SD}$ & $113.2 \pm 0.7$ & $149.5 \pm 0.9$ & $174.7 \pm 1.1$ & $254.3 \pm 1.6$ \\
 & $\sigma_\text{DD}$ & $2.138 \pm 0.014$ & $3.081 \pm 0.020$ & $3.643 \pm 0.023$ & $6.87 \pm 0.04$ \\\cline{2-6}
 & & \multicolumn{4}{c}{Fiducial} \\\cline{2-6}
 & $\sigma_\text{ND}^\text{fid.}$ & $2497 \pm 11$ & $3319 \pm 15$ & $3683 \pm 16$ & $5480 \pm 25$ \\
 & $\sigma_\text{SD}^\text{fid.}$ & $52.71 \pm 0.34$ & $61.7 \pm 0.4$ & $68.8 \pm 0.4$ & $84.4 \pm 0.6$ \\
 & $\sigma_\text{DD}^\text{fid.}$ & $0.784 \pm 0.005$ & $1.068 \pm 0.007$ & $1.229 \pm 0.008$ & $2.076 \pm 0.014$ \\
 
\bottomrule
\end{tabular}
\caption{Cross sections of non-diffractive (ND), single-diffractive (SD), and double-diffractive (DD) $Z$- and $W$-boson production with decays to muon-type leptons. The fiducial $Z$ selection requires both muons to satisfy $p_T^\mu>25~\mathrm{GeV}$ and $|\eta^\mu|<2.5$. The fiducial $W$ selection requires $p_T^\mu>25~\mathrm{GeV}$, $|\eta^\mu|<2.5$, and $p_T^\nu>25~\mathrm{GeV}$. No additional dimuon-mass or transverse-mass requirement is applied. Inclusive uncertainties are the numerical integration uncertainties stored by \textsc{Herwig}; fiducial uncertainties additionally include the finite-sample acceptance uncertainty in quadrature. DD production is listed for completeness but is excluded from the precision fits.}
\label{tab:nominal_yields}
\end{table}